\documentclass[a4paper,12pt]{article}

\usepackage[dvips]{graphicx}
\usepackage{pstricks}
\usepackage{bm}

\usepackage{graphicx}% Include figure files

\newcommand{\be}{\begin{equation}}
\newcommand{\ee}{\end{equation}}

\newcommand{\ka}{\kappa}

\def\beq{\begin{equation}}
\def\eeq{\end{equation}}

\def\al{\alpha}
\def\bt{\beta}
\def\Ga{\Gamma}

\def\de{\delta}
\def\De{\Delta}

\def\ka{\kappa}

\def\te{\theta}

\def\La{\Lambda}
\def\lam{\lambda}

\def\om{\omega}
\def\ep{\epsilon}

\def\sq{\sqrt}

\def\l{\left (}
\def\r{\right )}

\def\fr{\frac}
\def\la{\label}
\def\hs{\hspace}
\def\vs{\vspace}

\def\ran{\rangle}
\def\lan{\langle}
\def\ov{\overline}
\def\tl{\tilde}
\def\tm{\times}
\def\lrar{\leftrightarrow}

\begin{document}

\begin{flushright}
OSU-HEP-07-01\\
May 30, 2007 \\
%hep-ph/
\end{flushright}

\vs{0.5cm}

\begin{center}
{\Large\bf

Predictive Model of Inverted Neutrino Mass Hierarchy\\
\vs{0.2cm}
and Resonant Leptogenesis
}
\end{center}

\vspace{0.5cm}
\begin{center}
{\large
{}~K.S. Babu\footnote{E-mail: babu@okstate.edu},{}~
Abdel G. Bachri\footnote{E-mail: abdel.bachri@okstate.edu},
{}~Zurab Tavartkiladze\footnote{E-mail: zurab.tavartkiladze@okstate.edu}
}
\vspace{0.5cm}

{\em Department of Physics, Oklahoma State University, Stillwater, OK 74078, USA }
\end{center}
%\vspace{0.6cm}

\begin{abstract}

%\dspace

We present a new realization of inverted neutrino mass hierarchy based on $S_3 \times
{\cal U}(1)$ flavor symmetry. In this scenario, the deviation of the solar oscillation angle from
$\pi/4$ is correlated with the  value of $\theta_{13}$, as they are both induced by a common mixing
angle in the charged lepton sector.
We find several interesting predictions: $\te_{13}\geq 0.13$, $\sin^2\te_{12}\geq 0.31$, $\sin^2\te_{23}\simeq 0.5$ and
$0\leq \cos \de \leq 0.7$ for the neutrino oscillation parameters and $0.01~{\rm eV}\stackrel{<}{_\sim }m_{\bt \bt}\stackrel{<}{_\sim } 0.02~{\rm eV}$  for the effective neutrino mass in neutrino-less double $\bt $-decay. We show that our scenario can also explain naturally the observed baryon asymmetry of the universe via resonant leptogenesis. The masses of the decaying right--handed neutrinos can be in the range $(10^3 - 10^7)$ GeV, which would avoid the generic gravitino problem of supersymmetric models.

\end{abstract}

\vs{0.7cm}

%\dspace

\newpage

\section{Introduction}\label{sec:intro}

A lot has been learned about the pattern of neutrino masses and mixings over the
past decade from atmospheric \cite{Fukuda:2000np} and solar \cite{Fukuda:2001nj,Eguchi:2002dm} neutrino oscillation experiments.
When these impressive results are supplemented by results from reactor  \cite{Eguchi:2002dm}-\cite{Boehm:2001ik} and accelerator
\cite{Michael:2006rx}  neutrino oscillation experiments, a comprehensive picture for neutrino masses begins
to emerge. A global analysis of these results gives rather precise determination of some of
the oscillation parameters \cite{Maltoni:2004ei}-\cite{Fogli:2005cq}:
$$
|\De m_{\rm atm}^2|=2.4 \cdot \l 1^{+0.21}_{-0.26}\r \tm 10^{-3}~{\rm eV}^2~,~~~
\sin^2\te_{23}=0.44 \cdot \l 1^{+0.41}_{-0.22}\r ~,
$$
$$
\De m_{\rm sol}^2=7.92 \cdot \l 1\pm 0.09\r \tm 10^{-5}~{\rm eV}^2~,~~~
\sin^2\te_{12}=0.314 \cdot \l 1^{+0.18}_{-0.15}\r ~,
$$
\beq
\theta_{13} \stackrel{<}{_\sim } 0.2  ~.
\la{atm-sol-data}
\eeq
While these results are impressive, there are still many important
unanswered questions. One issue is the sign of $\De m_{\rm atm}^2=m^2_3-m^2_2$
which is presently unknown.  This is directly linked to nature of neutrino
mass hierarchy. A positive sign of $\De m_{\rm atm}^2$ would indicate normal hierarchy ($m_1 < m_2 < m_3$)
while a negative sign would correspond to an inverted mass hierarchy ($m_2 \stackrel{>}{_\sim } m_1 > m_3$).
Another issue is the value of the leptonic mixing angle $\theta_{13}$, which currently is
only bounded from above.  A third issue is whether CP is violated in neutrino oscillations, which is
possible (with $\te_{13}\neq 0$) if the phase angle $\delta$ in the MNS matrix is non--zero.
Forthcoming long baseline
experiments \cite{Michael:2006rx}, NO$\nu $A \cite{Ayres:2004js}, T2K \cite{Ishitsuka:2005qi,Hagiwara:2005pe}
and reactor experiments  double CHOOZ and Daya Bay will explore some or all these fundamental questions.
Answers to these have the potential for revealing the underlying symmetries of nature.

While there exists in the literature a large number of theoretical
models for  normal neutrino mass hierarchy, such is not the case with inverted
hierarchy. A large number of models for inverted hierarchy based on
symmetries \cite{Petcov:1982ya}-\cite{Babu:2002ex} that were proposed a few years ago are now
excluded by the solar and Kamland data, which proved that $\theta_{12}$ is significantly away
from the maximal value of $\pi/4$ predicted by most of these models. As a result, there is a dearth of
viable inverted neutrino mass hierarchy models. In this paper, we
attempt to take a step towards  remedying this situation.

Here we suggest a class of models for inverted neutrino mass hierarchy
based on $S_3 \times {\cal U}(1)$ flavor symmetry.  $S_3$ is the non-Abelian group
generated by the permutation of three objects, while the ${\cal U}(1)$ is used for explaining the
mass hierarchy of the leptons.  This ${\cal U}(1)$ symmetry is naturally identified with the anomalous
${\cal U}(1)$ of string origin.  In our construction, the $S_3$ permutation symmetry is broken down
to an approximate  $S_2$ in the neutrino sector, whereas it is broken completely in the charged lepton
sector.  Such a setup enables us to realize effectively a $\nu_\mu \leftrightarrow \nu_\tau$
interchange symmetry in the neutrino sector (desirable for maximal mixing in atmospheric neutrino oscillations), while
having non-degenerate charged leptons.  The ${\cal U}(1)$ symmetry acts as
leptonic $L_e-L_{\mu}-L_{\tau}$ symmetry, which is also desirable for an inverted neutrino
mass spectrum.  The breaking of $S_2$ symmetry in the charged lepton sector enables us to obtain
$\theta_{12}$ significantly different from $\pi/4$.

Interestingly, we find that the amount of deviation of $\theta_{12}$
from $\pi/4$ is determined by $\theta_{13}$ through the
relation
\begin{equation}
\sin^2\te_{12}
\simeq \fr{1}{2}-\tan \te_{13}\cos \de ~.
\la{prediction12}
\end{equation}
When compared with the neutrino data, the relation (\ref{prediction12}) implies the constraints (see Fig. 1):
\beq
\te_{13}\geq
0.13~,~~~~~~~~~|\de |\leq 0.75 \hs{0.2cm}(\simeq 43^o)~.
\la{bounds-te13-de}
\eeq
At the same time, the model gives
\beq
\sin^2\te_{23}\simeq \fr{1}{2}(1-\tan^2\theta_{13})~,
\la{prediction23}
\eeq
which is very close to 1/2.
These predictions will be tested in forthcoming experiments.
Somewhat similar relations have been obtained in scenarios with `quark-lepton complementarity'
\cite{Giunti:2002ye}-\cite{Hochmuth:2006xn}
by postulating the relations $\te_{12}+\te_c\approx \pi /4$, $\te_{23}+V_{cb}\approx \pi /4$ ($\te_c$ is the Cabibbo angle).
In our approach the leptonic mixing angles are inter-related by symmetries without involving the quark sector.
Furthermore, we are able to derive the relations (\ref{prediction12})-(\ref{prediction23}) from flavor symmetries
(see section \ref{sec:u1z3model}).

Our models have the right ingredients to generate the observed baryon asymmetry of the universe
via resonant leptogenesis. The ${\cal U}(1)$ symmetry which acts on leptons as
$L_e-L_{\mu}-L_{\tau}$ symmetry
guarantee that two right--handed neutrinos that we use for see saw mechanism are quasi-degenerate. This feature leads to a
resonant enhancement in the leptonic
CP asymmetry, which in turn admits low right--handed neutrino masses, as low as few TeV.
With such light right--handed neutrinos (RHN) generating lepton asymmetry, there is no
cosmological gravitino problem when these models are supersymmetrized.

The class of neutrino mass models and leptogenesis scenario that we present here will work well in
both  supersymmetric and non-supersymmetric contexts.
However, since low energy SUSY has strong phenomenological and theoretical
motivations, we shall adopt  the supersymmetric framework for our explicit constructions.

\section{Predictive Framework for Neutrino Masses and Mixings}\label{sec:neutrino}

In order to build inverted hierarchical neutrino mass matrices which are predictive and which lead to successful neutrino oscillations, it
is enough to introduce two right--handed neutrino states $N_{1, 2}$.
Then the superpotential relevant for neutrino masses is
\beq
W_{\nu }=l^TY_{\nu }Nh_u-\fr{1}{2}N^TM_NN~,
\la{Wnu}
\eeq
where $h_u$ denotes the up--type Higgs doublet superfield, while $Y_{\nu }$ and $M_N$ are $3\tm 2$ Dirac
Yukawa matrix and $2\tm 2$ Majorana mass matrix respectively.
Their structures can be completely determined by flavor symmetries. In order to have predictive models of
inverted hierarchy, the $L_e-L_{\mu }-L_{\tau }\equiv {\bf L}$ symmetry can be used
\cite{Petcov:1982ya}-\cite{Altarelli:2005pj}. This symmetry
naturally gives rise to large $\te_{23}$ and maximal $\te_{12}$ angles. At the same time, the mixing angle
$\te_{13}$ will be zero. In order to accommodate the  solar neutrino oscillations,
the ${\bf L}$-symmetry must be broken.
The pattern of ${\bf L}$-symmetry breaking will determine the relations and predictions for neutrino masses and
mixings. As a starting point, in the neutrino sector let us impose $\mu -\tau$ interchange symmetry $S_2$: $l_2\lrar l_3$, which will lead to maximal $\nu_{\mu }-\nu_{\tau }$ mixing, consistent with atmospheric neutrino data.

The leptonic mixing angles will receive contributions from both the neutrino sector and the charged lepton
sector.  As an initial attempt let us assume that the charged lepton mass matrix is diagonal.  We will
elaborate on altering this assumption in the next subsection.

For completeness, we will start with general couplings respecting the $S_2$ symmetry. Therefore, we have
\beq
\begin{array}{ccc}
 & {\begin{array}{cc}
\hs{-1.2cm}N_1 & N_2
\end{array}}\\ \vspace{0.5mm}
Y_{\nu }=
\begin{array}{c}
l_1\vs{0.1cm} \\ l_2\vs{0.1cm} \\ l_3
 \end{array}\!\!\!\!\!\hs{-0.2cm} &{\left(\begin{array}{cc}

 \al  &~0
 \vs{-0.5cm}
\\
&
\\
\bt'&~\bt
 \vs{-0.5cm}
 \\
 &
\\
 \bt'& ~\bt

\end{array}\right)}~,~~~~~~~
\end{array}
\begin{array}{cc}
 & {\begin{array}{cc}
\hs{-0.8cm}N_1 &\hs{0.2cm} N_2
\end{array}}\\ \vspace{1mm}
M_N=
\begin{array}{c}
N_1 \\ N_2
 \end{array}\!\!\!\!\!\hs{-0.2cm} &{\left(\begin{array}{cc}

 -\de_N &~~~1
\\
 ~1&~-\de_N^{\hs{0.3mm}'}

\end{array}\right)M}~.
\end{array}  \!\!
\label{Ynu-MN}
\eeq
Note that setting $(1,2)$ element of $Y_{\nu }$ to zero can be done without loss of generality -
by a
proper redefinition of $N_{1,2}$ states. The couplings $\al , \bt $ and $(1,2), (2,1)$ entries in $M_N$ respect
${\bf L}$ symmetry, while the couplings $\bt', \de_N$ and $\de_N^{\hs{0.3mm}'}$ violate it. Therefore, it is natural to assume that
$|\bt'|\ll |\al |, |\bt |$, and $|\de_N|, |\de_N^{\hs{0.3mm}'}|\ll 1$. Furthermore, by proper field redefinitions all
couplings in $Y_{\nu }$ can be taken to be real. Upon these redefinitions $\de_N$ and $\de_N^{\hs{0.3mm}'}$ entries in $M_N$ will be complex.

Integration of the heavy $N_{1,2}$ states leads to the following $3\tm 3$ light neutrino mass matrix:
\begin{equation}
\begin{array}{ccc}
 & {\begin{array}{ccc}
 & &
\end{array}}\\ \vspace{1mm}

\begin{array}{c}
~\\ ~ \\~
 \end{array}\!\!\!\!\!\hs{-0.1cm}m_{\nu }= \hs{-0.3cm}&{\left(\begin{array}{ccc}

 2\de_{\nu }^{\hs{0.3mm}'} & \sq{2} &\sq{2}
\\
\sq{2}&\de_{\nu } &\de_{\nu }
 \\
\sq{2}&\de_{\nu } & \de_{\nu }
\end{array}\right)\fr{m}{2}}~,
\end{array}  \!\!  ~~~~~
\label{mnu}
\end{equation}
where
$$
m=\fr{\lan h_u^0\ran^2}{M(1-\de_N\de_N^{\hs{0.3mm}'})}\sq{2}\al \l \bt +\bt'\de_N^{\hs{0.3mm}'}\r ~,
$$
\beq
\de_{\nu }=\fr{\sq{2}}{\al }\fr{2\bt \bt'+\bt^2\de_N+(\bt')^2\de_N^{\hs{0.3mm}'}}{\bt +\bt'\de_N^{\hs{0.3mm}'}}~,~~~~~
\de_{\nu }^{\hs{0.3mm}'}=\fr{\al }{\sq{2}}\fr{\de_N^{\hs{0.3mm}'}}{\bt +\bt'\de_N^{\hs{0.3mm}'}}~.
\la{m-denus}
\eeq
The entries $\de_{\nu }$, $\de_{\nu }^{\hs{0.3mm}'}$ in (\ref{mnu}) are proportional to the ${\bf L}$-symmetry breaking couplings
and therefore one naturally expects $|\de_{\nu }|, |\de_{\nu }^{\hs{0.3mm}'}|\ll 1$. These small entries
are responsible for  $\De m_{\rm sol}^2\neq 0$, i.e. for the solar neutrino oscillation.
The neutrino mass matrix is diagonalized by unitary transformation
$U_{\nu }^Tm_{\nu }U_{\nu }={\rm Diag}\l m_1, m_2, 0\r $, were $U_{\nu }=U_{23}U_{12}$ with
\begin{equation}
\begin{array}{ccc}
 & {\begin{array}{ccc}
\hs{-1.2cm} &
\end{array}}\\ \vspace{0.5mm}
U_{23}=
\begin{array}{c}
 \\  \\
 \end{array}\!\!\!\!\!\hs{-0.2cm} &{\left(\begin{array}{ccc}

 1 &~0 & ~0
 \vs{-0.5cm}
\\
&
\\
0& \fr{1}{\sq{2}}  & -\fr{1}{\sq{2}}
 \vs{-0.5cm}
 \\
 & &
\\
 0& ~\fr{1}{\sq{2}}  & ~\fr{1}{\sq{2}}

\end{array}\right)}~,~~~~~~~
\end{array}
\begin{array}{cc}
 & {\begin{array}{cc}
\hs{-0.8cm}&\hs{0.2cm}
\end{array}}\\ \vspace{1mm}
U_{12}\simeq
\begin{array}{c}
\\
 \end{array}\!\!\!\!\!\hs{-0.2cm} &{\left(\begin{array}{ccc}

 \bar c &~~-\bar se^{i\rho }  & 0
\\
\bar se^{-i\rho } &~\bar c &0
 \\
  0& 0 & 1

\end{array}\right)}~,
\end{array}  \!\!
\label{Unu}
\end{equation}
where $\bar c=\cos \bar \te$, $\bar s=\sin \bar \te$ and
\beq
\tan \bar \te \simeq 1\pm \fr{1}{2}\ka ~,~~~~~\ka=
\fr{|\de_{\nu }|^2-|\de_{\nu }^{\hs{0.3mm}'}|^2}{|\de_{\nu }^*+\de_{\nu }^{\hs{0.3mm}'}|}~.
\la{tan-barte}
\eeq
The phase $\rho $ is determined from the relation
\beq
|\de_{\nu }|\sin (\om_{\nu }-\rho )=|\de_{\nu }^{\hs{0.3mm}'}|\sin (\om_{\nu }^{\hs{0.3mm}'}+\rho )~,~~~
\om_{\nu }={\rm Arg}(\de_{\nu })~,~~\om_{\nu }^{\hs{0.3mm}'}={\rm Arg}(\de_{\nu }^{\hs{0.3mm}'})~,
\la{eq-rho}
\eeq
and should be taken such that
\beq
|\de_{\nu }|\cos (\om_{\nu }-\rho )+|\de_{\nu }^{\hs{0.3mm}'}|\cos (\om_{\nu }^{\hs{0.3mm}'}+\rho )<0~.
\la{cond-rho}
\eeq
This condition ensures $\De m_{\rm sol}^2=m_2^2-m_1^2>0$ needed for solar neutrino oscillations. For $\De m_{\rm atm}^2$ and
the ratio $\De m_{\rm sol}^2/|\De m_{\rm atm}^2|$ we get
\beq
|\De m_{\rm atm}^2|\simeq |m|^2~,~~~
\fr{\De m_{\rm sol}^2}{|\De m_{\rm atm}^2|}\simeq
-2\l |\de_{\nu }|\cos (\om_{\nu }-\rho )+|\de_{\nu }^{\hs{0.3mm}'}|\cos (\om_{\nu }^{\hs{0.3mm}'}+\rho )\r
=2\left | \de_{\nu }^*+\de_{\nu }^{\hs{0.3mm}'}\right |~.
\la{sol-atm-ratio}
\eeq

With no contribution from the charged lepton sector, the leptonic mixing matrix is $U_{\nu }$. From
(\ref{Unu}), (\ref{tan-barte}) for the solar mixing angle we will have $\sin^2\te_{12}=\fr{1}{2}\pm \fr{\ka }{4}$.
In order to be compatible with experimental data one needs $\ka \approx 0.7$. On the other hand with $|\de_{\nu }|\sim |\de_{\nu }^{\hs{0.3mm}'}|$
and no specific phase alignment from (\ref{sol-atm-ratio}) we estimate $|\de_{\nu }|\sim |\de_{\nu }^{\hs{0.3mm}'}|\sim 10^{-2}$. Thus
we get the expected value $\ka \sim 10^{-2}$, but with the $\te_{12}$ mixing angle nearly maximal, which is incompatible with experiments.
This picture remains unchanged with the inclusion of renormalization group effects.
Therefore, we learn that it is hard to accommodate  the neutrino data in simple minded inverted hierarchical neutrino mass scenario.
In order for  the scenario to be compatible with the experimental data we need simultaneously
\beq
\left | \de_{\nu }^*+\de_{\nu }^{\hs{0.3mm}'}\right |=\fr{\De m_{\rm sol}^2}{2|\De m_{\rm atm}^2|}\simeq 0.016~~~~~{\rm and}~~~~~~
\fr{|\de_{\nu }|^2-|\de_{\nu }^{\hs{0.3mm}'}|^2}{|\de_{\nu }^*+\de_{\nu }^{\hs{0.3mm}'}|}=\mp (0.52-0.92)~.
\la{simult-conds}
\eeq
Therefore, one combination of $\de_{\nu }$ and $\de_{\nu }^{\hs{0.3mm}'}$ must be $\sim 50$-times larger than the
other.
This is indeed unnatural and no explanation for these conditions is provided at this stage. To make this
point more clear let's consider the case with $\de_{\nu }=0$. In this case from (\ref{sol-atm-ratio}) we have
$|\de_{\nu }^{\hs{0.3mm}'}|\simeq 0.016$. Using this in (\ref{tan-barte}) we obtain $\sin^2\te_{12}\geq 0.496$,
which is excluded by the solar neutrino data (\ref{atm-sol-data}).

Summarizing, although the conditions in (\ref{simult-conds}) can be satisfied, it remains a challenge to have a
natural explanation of needed hierarchies. This is a shortcoming of the minimal scenario. Below we present
a possible solution to this conundrum  which looks attractive and maintains predictive power without fine tuning
by making use of mixing in the charged lepton sector.

 \subsection{Improved $\te_{12}$ with $\te_{13}\neq 0$}\label{sec:imprmodel}

Let us now include the charged lepton sector in our studies. The relevant superpotential is
\beq
W_{e}=l^TY_Ee^ch_d~,
\la{W_e}
\eeq
where $Y_E$ is $3\tm 3$ matrix in the family space. In general, $Y_E$ has off--diagonal entries.
Being so, $Y_E$ will induce contributions to
the leptonic mixing matrix. We will use this contribution in order to fix the value of $\te_{12}$ mixing angle.
It is desirable to do this in such a way that some predictivity is maintained. As it turns out, the texture
\begin{equation}
\begin{array}{ccc}
 & {\begin{array}{ccc}
 & &
\end{array}}\\ \vspace{1mm}

\begin{array}{c}
~\\ ~ \\~
 \end{array}\!\!\!\!\!\hs{-0.1cm}Y_E= \hs{-0.3cm}&{\left(\begin{array}{ccc}

 0 & a' &0
\\
a&\lam_{\mu } &0
 \\
 0&0 & \lam_{\tau }
\end{array}\right)}~,
\end{array}  \!\!  ~~~~~
\label{YE}
\end{equation}
gives interesting predictions. In the structure (\ref{YE}) there is only one irremovable complex
phase and we leave it in (1,2) entry. Thus, we make the parametrization $a'=\lam_{\mu }\te_ee^{i\om }$, while all the
remaining entries can be taken to be real.
Diagonalizing  $Y_EY_E^{\dagger }$,
 namely,
$U_eY_EY_E^{\dagger }U_e^{\dagger }=\l Y_E^{\rm diag}\r^2$, it is easy to see that
\begin{equation}
\begin{array}{ccc}
 & {\begin{array}{ccc}
 & &
\end{array}}\\ \vspace{1mm}

\begin{array}{c}
~\\ ~ \\~
 \end{array}\!\!\!\!\!\hs{-0.1cm}U_e= \hs{-0.3cm}&{\left(\begin{array}{ccc}

 c & se^{i\om } &0
\\
-se^{-i\om }&c&0
 \\
 0&0 & 1
\end{array}\right)}~,
\end{array}  \!\!  ~~~~~
\label{Ue}
\end{equation}
where $c\equiv \cos t$, $s\equiv \sin t$ and $\tan t=-\te_e$~.
%
%  Figure 1
%
\begin{figure}[t]
\centerline{
\includegraphics*[height=5in,viewport=90 467 417 782]{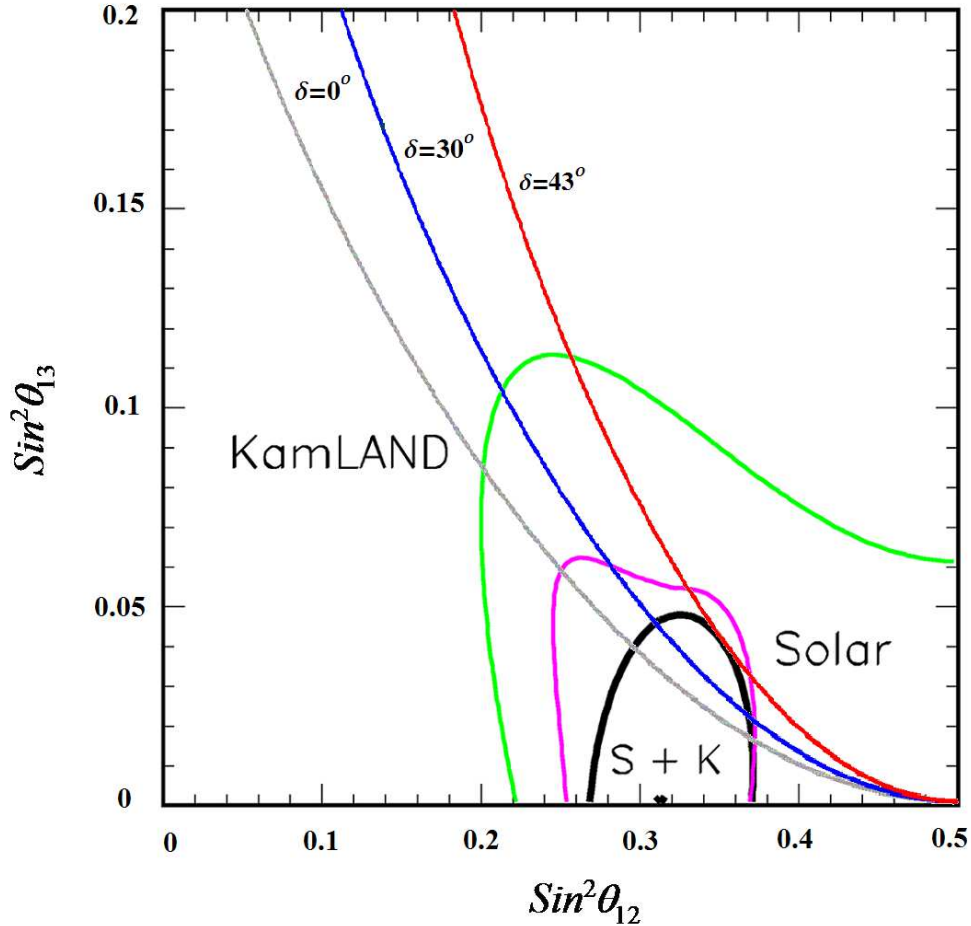}
%\put(-262,316){\small $|~|$}
%\put(-220,295){\small $|~|$}
%\put(-178,283){\small $|~|$}
}
\vspace*{0.2cm} \caption{Correlation
between $\theta_{12}$ and $\theta_{13}$ taken from Fogli {\it et al} of Ref. \cite{Fogli:2005cq}.
Three `sloped' curves correspond to $\theta_{12}-\theta_{13}$ dependence
(for three different absolute values of CP phase $\de $) obtained from our model
according to Eq. (\ref{pred-12-23}).
 \label{fig1}}
\end{figure}
Finally, the leptonic mixing matrix takes the form
\beq
U^l=U_e^*U_{\nu }~,
\la{Ulept}
\eeq
where $U_{\nu }=U_{23}U_{12}$ can be derived from Eq. (\ref{Unu}).
Therefore, for the corresponding mixing elements we get
\beq
U^l_{e3}=-\fr{s}{\sq{2}}e^{-i(\om +\rho )}~,~~
|U^l_{e2}|=\fr{1}{\sq{2}}\left | c-\fr{s}{\sq{2}}e^{-i(\om +\rho )}\right |~,~~
|U^l_{\mu 3}|=\fr{c}{\sq{2}}~.
\la{Ulept-elements}
\eeq
Comparing these with those written in the standard parametrization of $U_{\rm MNS}$ we obtain the relations
\beq
s_{13}=-\fr{s}{\sq{2}}~,~~~~\om +\rho =\de +\pi ~,
\la{s13}
\eeq
\beq
s_{12}c_{13}=|U^l_{e2}|~,~~~s_{23}c_{13}=|U^l_{\mu 3}|~.
\la{s12-s23}
\eeq
Using (\ref{s13}) and (\ref{Ulept-elements}) in (\ref{s12-s23})  we arrive at the following predictions:
$$
\sin^2\te_{12}=\fr{1}{2}-\sq{1-\tan^2\te_{13}}\tan \te_{13}\cos \de ~,
$$
\beq
\sin^2\te_{23}=\fr{1}{2}\l 1-\tan^2\te_{13}\r ~.
\la{pred-12-23}
\eeq
Since the CHOOZ results require $s_{13}\stackrel{<}{_\sim }0.2$, the first relation in (\ref{pred-12-23}), with the help of
the solar neutrino data provides an upper bound for absolute value of the CP violating phase:
$|\de |\stackrel{<}{_\sim }|\de |_{\rm max}\approx 0.84 (\simeq48^o)$.
However, this estimate ignores the dependence of $\te_{12}$ on the value of $\te_{13}$ in the neutrino
oscillation data.
Having $\te_{13}\neq 0$, this dependence shows up because one deals with three flavor oscillations.
This has been analyzed in Ref. \cite{Fogli:2005cq}. We show the results in  Fig. 1 (borrowed from  Ref. \cite{Fogli:2005cq}) along with the constraints arising
from our model.
We have shown three curves corresponding
to (\ref{pred-12-23}) for different values of $|\de |$. Now we see that maximal allowed value for $|\de |$ is $|\de |_{\rm max}\simeq 0.75 (\simeq 43^o)$.
Moreover, for a given $\de $ we predict the allowed range for
$\te_{13}$. In all cases the values are such that these relations can be tested in the near future. An interesting result from our scenario is that we  obtain
lower and upper bounds for $\te_{13}$ and  $|\de |$ respectively
\beq
\te_{13}\geq 0.13~,~~~~~~~~~~|\de |\leq 0.75 {\hs{0.1cm}}(\simeq 43^o)~.
\la{bounds-te13-de1}
\eeq

\begin{figure}[t]
\centerline{
\includegraphics*[height=3.95in,width=5.7in,viewport=84 88 386 265]{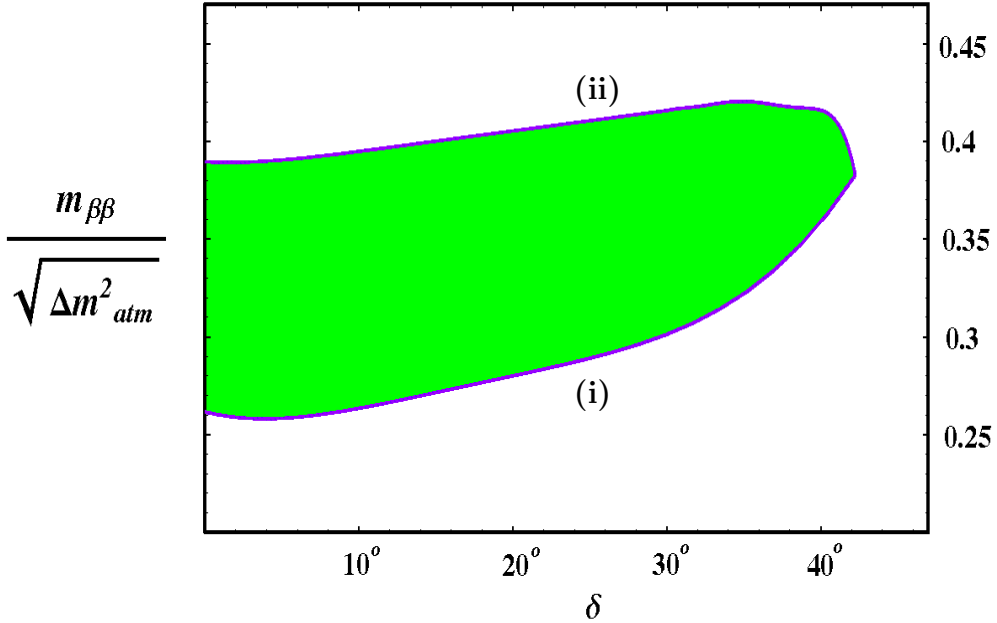}
\put(-175,100){\bf (i)}
\put(-175,215){\bf (ii)}
%\put(-175,20){\bf $|~~|$}
}
\caption{Curves ({\bf i}) and ({\bf ii})  respectively show the dependence of {\bf \large $\fr{m_{\bt \bt }}{\sq{\De m_{\rm atm }^2}}$}'s
low and upper bounds  on absolute value of CP violating phase $\de $. The shaded region
 corresponds to  values of $m_{\bt \bt}$ and $|\de |$ realized within our model.}
\label{fig:2}
\end{figure}

Finally, the neutrino-less double $\bt $-decay parameter in this scenario is given by
\beq
 m_{\bt \bt } \simeq 2\sq{\De m_{\rm atm}^2} \tan \te_{13}\fr{\sq{1-\tan^2\te_{13}}}{\sq{1+\tan^2\te_{13}}}~.
\la{2btdecay}
\eeq
We have neglected the small contribution (of order $\Delta m^2_{\rm solar}/\sq{\Delta m^2_{\rm atm}}$) arising
from the neutrino mass matrix diagonalization.
Since the value of $\te_{13}$ is experimentally constrained ($\stackrel{<}{_\sim }0.2$), to a good approximation we
have $m_{\bt \bt } \approx 2\sq{\De m_{\rm atm}^2}\tan \te_{13}$. Using this result and the atmospheric
neutrino data (\ref{atm-sol-data}) we find $m_{\bt \bt } \stackrel{<}{_\sim } 0.02$~eV.
Knowledge of $\te_{13}$-dependence on $|\de |$ (see Fig. 1) allows us to make
more accurate estimates for the range of $ m_{\bt \bt } $ for each given value of $|\de |$. The dependence of   $m_{\bt \bt }$  on $|\de |$ is given
in Fig. 2. We have produced this graph with the predictive relations (\ref{pred-12-23}), (\ref{2btdecay}) using the neutrino data \cite{Fogli:2005cq}.
Combining these results we arrive at
\beq
0.011~{\rm eV}\stackrel{<}{_\sim } m_{\bt \bt } \stackrel{<}{_\sim }0.022~{\rm eV}.
\la{range-mee}
\eeq
We see that the predicted range, depending on the value of $|\de |$, is quite narrow.  Future measurements of CP violating
phase $\de $ together with a discovery of the neutrino-less double $\bt $-decay
will be another test for the inverted hierarchical scenario presented here.

\section{Resonant Leptogenesis}\label{sec:leptogenesis}

Neutrino mass models with heavy right--handed neutrinos provide an attractive and natural
framework for explaining the observed baryon asymmetry of the universe through thermal leptogenesis \cite{Fukugita:1986hr}.
This mechanism takes advantage of the out-of-equilibrium decay of lightest right--handed neutrino(s) into
leptons and the Higgs boson.
In the scenario with hierarchical RHNs, a lower bound on the mass of decaying RHN
has been derived: $M_{N_1} \geq 10^9$~GeV \cite{Davidson:2002qv,Buchmuller:2002rq} (under some
not so unreasonable assumptions\footnote{See ref. \cite{Raidal:2004vt} for scenarios
which violate this limit with hierarchical RHN masses.}). The reheating temperature cannot be much below
the mass of ${N_1}$. In low energy SUSY models (with $m_{3/2}\sim 1$~TeV) this is in conflict with the upper bound on reheating
temperature obtained from the gravitino abundance \cite{Khlopov:1984pf}-\cite{Olive:2006uw}. This conflict can be naturally avoided in
the scenario of `resonant leptogenesis' \cite{Flanz:1996fb}-\cite{Pilaftsis:2003gt}. Due to the quasi-degeneracy in mass
of the RHN states, the
needed CP asymmetry can be  generated even if the right--handed neutrino mass is lower than $10^9$ GeV.

Our model of inverted hierarchical neutrinos involves two quasi-degenerate RHN states and has
all the needed ingredients for successful resonant leptogenesis.  This makes the scenario
attractive from a cosmological viewpoint as well. Now we  present a detailed study
of the resonant leptogenesis phenomenon in our scenario.

 The CP asymmetry is created by resonant out of equilibrium decays of $N_1, N_2$
and is given by \cite{Pilaftsis:1997jf,Pilaftsis:2003gt}
\beq
\ep_1=\fr{{\rm Im}(\hat{Y}_{\nu }^{\dagger}\hat{Y}_{\nu })_{21}^2}
{(\hat{Y}_{\nu }^{\dagger}\hat{Y}_{\nu })_{11}(\hat{Y}_{\nu }^{\dagger}\hat{Y}_{\nu })_{22}}
\fr{\l M_2^2-M_1^2\r M_1\Ga_2}{\l M_2^2-M_1^2\r^2+M_1^2\Ga_2^2}~,
\la{res-lept-asym}
\eeq
with a similar expression for $\ep_2$. The asymmetries\footnote{Here we use asymmetries averaged in relatively large time interval.
The `memory' effects \cite{De Simone:2007rw} might cause changes in some cases.}
 $\ep_1$ and $\ep_2$ correspond to the decays of $N_1$ and $N_2$ respectively. Here $M_1, M_2$ are the mass eigenvalues of the matrix
$M_N$  in (\ref{Ynu-MN}), while $\hat{Y}_{\nu }=Y_{\nu }U_N$ is the Dirac Yukawa matrix in a basis where RHN
mass matrix is diagonal. The tree--level decay width of $N_i$ is given as $\Ga_i=(\hat{Y}_{\nu }^{\dagger}\hat{Y}_{\nu })_{ii}M_i/(8\pi )$.
The expression (\ref{res-lept-asym}) deals with the regime $M_2-M_1\sim \Ga_{1,2}/2$  (relevant for our studies)
consistently and has the correct behavior in the limit $M_1\to M_2$ \cite{Pilaftsis:1997jf,Pilaftsis:2003gt}.
{}From (\ref{Ynu-MN}) we have
\begin{equation}
\begin{array}{cc}
 & {\begin{array}{cc}
\hs{-0.8cm}&\hs{0.2cm}
\end{array}}\\ \vspace{1mm}
U_N^TM_NU_N={\rm Diag}\l M_1 , M_2\r ~,~~~~~~U_N\simeq
\begin{array}{c}
\\
 \end{array}\!\!\!\!\!\hs{-0.2cm} &{\fr{1}{\sq{2}}\left(\begin{array}{ccc}

 1 &~~-e^{ir }
\\
 e^{- ir} &~1

\end{array}\right)}~,
\end{array}  \!\!
\label{UN}
\end{equation}
with
\beq
M_2^2-M_1^2=2M^2\left | \de_N^*+\de_N^{\hs{0.3mm}'}\right |~,~~~~~~
\tan r=\fr{{\rm Im}\l \de_N-\de_N^{\hs{0.3mm}'}\r }{{\rm Re}\l \de_N+\de_N^{\hs{0.3mm}'}\r }~.
\la{sqdif-r}
\eeq
Introducing the notations
\beq
\fr{\al }{\bt }=x~,~~~~~~\fr{\bt'}{\bt }=x'~,
\la{xx1}
\eeq
we can write down the appropriate matrix elements needed for the calculation of leptonic asymmetry:
$$
(\hat{Y}_{\nu }^{\dagger}\hat{Y}_{\nu })_{11}=\fr{1}{2}\bt^2\l 2+x^2+2(x')^2+4xx'\cos r\r ~,
$$
$$
(\hat{Y}_{\nu }^{\dagger}\hat{Y}_{\nu })_{22}=\fr{1}{2}\bt^2\l 2+x^2+2(x')^2-4xx'\cos r\r ~,
$$
\beq
{\rm Im}(\hat{Y}_{\nu }^{\dagger}\hat{Y}_{\nu })_{21}^2=-\fr{1}{4}\bt^4\l 2-x^2-2(x')^2+4xx'\cos r\r^2\sin 2r ~.
\la{YY-elements}
\eeq
In terms of these entries the CP asymmetries are give by

\beq
\ep_1=\fr{{\rm Im}(\hat{Y}_{\nu }^{\dagger}\hat{Y}_{\nu })_{21}^2}{(\hat{Y}_{\nu }^{\dagger}\hat{Y}_{\nu })_{11}}
\fr{|\de_N^*+\de_N^{\hs{0.3mm}'}|}{16\pi |\de_N^*+\de_N^{\hs{0.3mm}'}|^2+(\hat{Y}_{\nu }^{\dagger}\hat{Y}_{\nu })_{22}^2/(16\pi )}~,~~~~~~~
\ep_2=-\ep_1(1\lrar 2)~.
\la{res-eps-12-byYY}
\eeq
Since we have five independent parameters, in general one should evaluate the lepton asymmetry as
a function of $x, x', |\de_N|, |\de_N^{\hs{0.3mm}'}|$
and $r$. Below we will demonstrate that resonant decays of $N_{1,2}$ can generate the needed CP asymmetry.

It turns out that for our purposes we will need $|\de_N^*+\de_N^{\hs{0.3mm}'}|\ll 1$. This, barring precise cancellation,
implies $|\de_N|, |\de_N^{\hs{0.3mm}'}|\ll 1$. From the symmetry viewpoint and also from further studies,  it turns out that
$\left | \fr{x'}{x}\right |\ll 1$ is a self consistent condition. Taking this condition and the results from the neutrino
sector into account, to a good approximation we have
\beq
\bt^2=\fr{\sq{\De m_{\rm atm}^2}M}{\sq{2}x\lan h_u^0\ran^2}~,~~~~
%\left |\fr{x'}{x}\right |=\fr{1}{4\sq{2}}\fr{\De m_{\rm sol}^2}{|\De m_{\rm atm}^2|}\simeq 6\cdot 10^{-3}~
\la{2conds}
\eeq
and
$$
\ep_1\simeq \ep_2\simeq \fr{{\rm Im}(\hat{Y}_{\nu }^{\dagger}\hat{Y}_{\nu })_{12}^2}{(\hat{Y}_{\nu }^{\dagger}\hat{Y}_{\nu })_{11}}
\fr{|\de_N^*+\de_N^{\hs{0.3mm}'}|}{16\pi |\de_N^*+\de_N^{\hs{0.3mm}'}|^2+(\hat{Y}_{\nu }^{\dagger}\hat{Y}_{\nu })_{11}^2/(16\pi )}\simeq
$$
\beq
-\fr{(2-x^2)^2}{2(2+x^2)} \bt^2\fr{|\de_N^*+\de_N^{\hs{0.3mm}'}|}
{16\pi |\de_N^*+\de_N^{\hs{0.3mm}'}|^2+(2+x^2)^2\bt^4/(64\pi )}\sin 2r ~,
\la{epses}
\eeq
where in the last expression we have ignored $x'$ contributions. This approximation is good
for all practical purposes.
The combination $|\de_N^*+\de_N^{\hs{0.3mm}'}|$ is a free parameter and since we are looking for a resonant regime, let us maximize the
expression in (\ref{epses}) with respect to this variable. The maximal CP asymmetry is achieved with
$|\de_N^*+\de_N^{\hs{0.3mm}'}|=(\hat{Y}_{\nu }^{\dagger}\hat{Y}_{\nu })_{11}/(16\pi )$.  Plugging this value back in (\ref{epses}) and
taking into account (\ref{YY-elements}), (\ref{2conds}) we arrive at
\beq
\bar{\ep }_1\simeq \bar{\ep }_2\simeq -\fr{(2-x^2)^2}{2(2+x^2)^2}\sin 2r~,
\la{epses-x1small}
\eeq
where $\bar{\ep }_{1,2}$ indicate the maximized expressions,
which do not depend on the scale of right--handed neutrinos.
We can take these masses as low as TeV! The expression in (\ref{epses-x1small})
reaches the maximal values for $x\ll 1$ and $x\gg 1$. However, the final value of $x$ will be fixed from the observed baryon asymmetry.
%
%
%          Resonant Plot
%
%

\begin{figure}[t]
\centerline{
\includegraphics*[height=5.5in,viewport=-25 387 637 820]{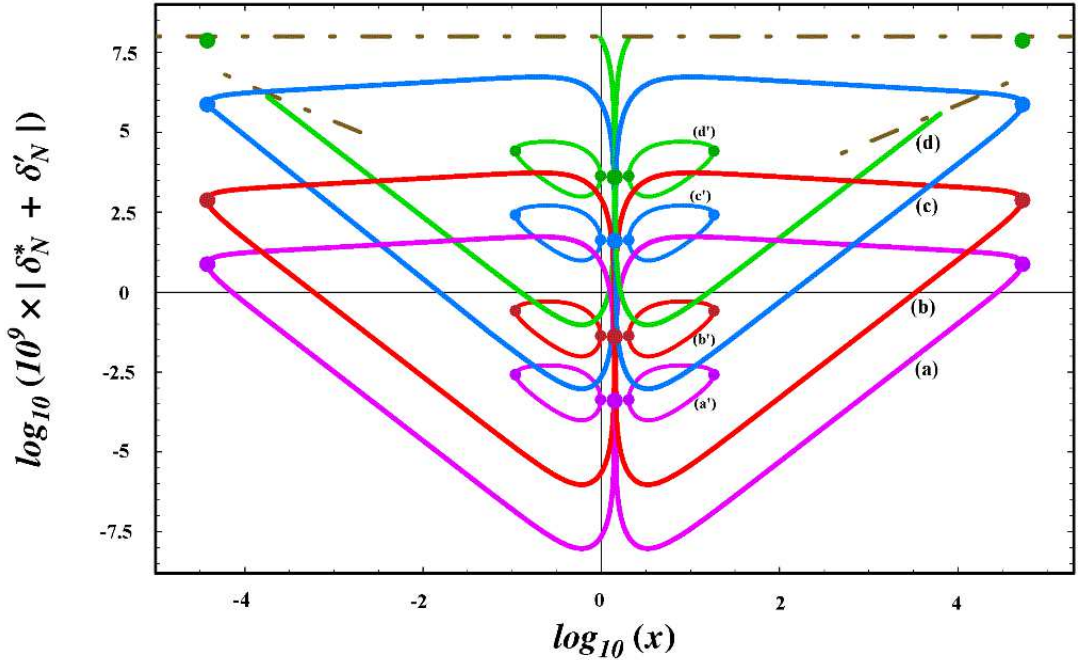}}
\vspace*{-3.2cm}
\caption{Resonant leptogenesis for inverted mass hierarchical neutrino scenario.
In all cases $\fr{n_B}{s}=9\tm 10^{-11}$ and $\tan \bt \simeq 2$. Curves {\bf (a)}, {\bf (b)},
{\bf (c)}, {\bf (d)} correspond respectively to the cases with
$M=\l 10^4 , 10^6 , 10^9 , 10^{11}\r $~GeV and $r=\pi /4$. The curves with primed labels correspond
to same  values of $M$, but with CP phase $r=5\cdot 10^{-5}$.
Bold dots stand for a maximized values of CP asymmetry [see Eq. (\ref{deg-factor})].
The `cut off' with horizontal dashed line reflects the requirement
$\left |\de_N^*+\de_N^{\hs{0.3mm}'}\right |\stackrel{<}{_\sim }0.1$. Two sloped dashed lines
restrict low parts of the `ovals' of $M=10^{11}$~GeV, insuring the Yukawa coupling
perturbativity.
\label{fig:3}}
\end{figure}
The lepton asymmetry is converted to the baryon asymmetry via sphaleron effects \cite{Kuzmin:1985mm}
and is given by
$\fr{n_B}{s}\simeq -1.48\cdot 10^{-3}(\ka_f^{(1)}\ep_1+\ka_f^{(2)}\ep_2)$, where $\ka_f^{(1,2)}$
are efficiency factors given approximately by \cite{Giudice:2003jh}
$$
\ka_f^{(1,2)}=\l \fr{3.3\cdot 10^{-3}~{\rm eV}}{\tl{m}_{1,2}}+\l \fr{\tl{m}_{1,2}}{0.55\cdot 10^{-3}~{\rm eV}}\r^{1.16}\r^{-1}~,~~~
$$
\beq
{\rm with}~~~\tl{m}_1=\fr{\lan h_u^0\ran^2}{M_1}(\hat{Y}_{\nu }^{\dagger}\hat{Y}_{\nu })_{11}~,~~
\tl{m}_2=\fr{\lan h_u^0\ran^2}{M_2}(\hat{Y}_{\nu }^{\dagger}\hat{Y}_{\nu })_{22}~.
\la{efic}
\eeq
In our model,  with $\left |\fr{x'}{x}\right |\ll 1$ we have
\beq
\tl{m}_1\simeq \tl{m}_2\simeq \fr{\sq{\De m_{\rm atm}^2}}{2\sq{2}x}(2+x^2)\simeq 0.017~{\rm eV}\tm \fr{2+x^2}{x}~.
\la{tlm12}
\eeq
This also gives $\ka_f^{(1)}\simeq \ka_f^{(2)}\equiv \ka_{f}$ and as a result we obtain
\beq
\left. \fr{n_B}{s}\right |_{\ep =\bar{\ep }}\simeq 1.48\cdot 10^{-3} \ka_{f}(x)\fr{(2-x^2)^2}{(2+x^2)^2}\sin 2r ~.
\la{nb-s}
\eeq
With $\sin 2r =1$ in order to reproduce the experimentally
observed value  $\l \fr{n_B}{s}\r^{\rm exp}=9\cdot 10^{-11}$
we have four possible choices of $x$: $x=3.8\cdot 10^{-5}$, $x=5.3\cdot 10^{4}$, $x=\sq{2}-0.0047$ or $x=\sq{2}+0.0047$. For  these  values of $x$ we have
respectively
$$
\left |\de_N^*+\de_N^{\hs{0.3mm}'}\right |_{\ep =\bar{\ep }}\simeq \fr{2+x^2}{32\sq{2}\pi x}\fr{\sq{\De m_{\rm atm}^2}M}{\lan h_u^0\ran^2}
\simeq
$$
\beq
\l 6\cdot 10^{-7}~, ~6\cdot 10^{-7}~,~ 3.2\cdot 10^{-11}~,~ 3.2\cdot 10^{-11}\r \tm \fr{1+\tan^2\bt }{\tan^2\bt }\fr{M}{{10^6\rm GeV}}
\la{deg-factor}
\eeq
(fixed from the condition of maximization). The MSSM parameter $\tan \bt $ should not be confused with Yukawa coupling in (\ref{2conds})).
Note that these results are obtained at the resonant regime  $|M_2-M_1|= \Ga_{1,2}/2$. If we are away from
this point, then the baryon asymmetry will be more suppressed and we will need to take different values of $x$.
In Fig. \ref{fig:3}  we show $\left |\de_N^*+\de_N^{\hs{0.3mm}'}\right |-x$ dependence corresponding to baryon asymmetry of $9\cdot 10^{-11}$. The curves are constructed with Eqs. (\ref{2conds}), (\ref{epses}).
We display different cases for different values of the mass $M$ and for two values of $CP$ violating phase $r$. For
smaller values of $r$ the `ovals' shrink indicating that there is less room in $\left |\de_N^*+\de_N^{\hs{0.3mm}'}\right |-x$
plane for generating the needed baryon asymmetry. We have limited ourselves to
$\left |\de_N^*+\de_N^{\hs{0.3mm}'}\right |\stackrel{<}{_\sim }0.1$. Above this value the degeneracy disappears and the validity
of our expression (\ref{res-lept-asym}) breaks down\footnote{There will be another contributions to the CP asymmetry, the vertex diagram, which
would be significant in the non-resonant case.}. Also, in this regime the inverted mass hierarchical neutrino scenario
becomes unnatural. The dashed horizontal line in Fig. \ref{fig:3} corresponds to this `cut--off'. This limits the cases with
larger masses [case (d) in Fig. \ref{fig:3}, of $M=10^{11}$~GeV]. The sloped dashed cut--off lines appear due to the requirement that the Yukawa couplings be perturbative ($\al , \bt \stackrel{<}{_\sim }1$). As one can see from (\ref{2conds}), for sufficiently large values of
$M$, with $x\gg 1$ or $x\ll 1$,  one of the Yukawa couplings becomes non-perturbative.

As we see, in some cases (especially for suppressed values of $r$) the degeneracy in mass between $N_1$ and $N_2$ states is required to be
very accurate, i.e. $\left |\de_N^*+\de_N^{\hs{0.3mm}'}\right |\ll 1$.
In section \ref{sec:u1z3model} we discuss the possibility for explaining this based on symmetries.

\section{Model with $S_3\tm {\cal U}(1)$ Symmetry}\label{sec:u1z3model}

In this section we present a concrete model which generates the needed textures for the charged lepton and the neutrino
mass matrices. The Lagrangian of the model is the most general under the symmetries of the model.
The model explains the hierarchies of the charged leptons, neutrinos and the lepton mixing angles.
Therefore, the relations (\ref{pred-12-23}) are derived as a consequence of symmetries.

The model presented here also blends in well with the
leptogenesis scenario investigated in the previous section.
In particular, the splitting between the masses of nearly degenerate heavy neutrinos have the right magnitude needed for
resonant leptogenesis.

We wish to have an
understanding of the hierarchies and the needed zero entries in the Dirac and Majorana neutrino couplings. Also, the
values of masses $M_{N_{1,2}}\simeq M\stackrel{<}{_\sim }10^8$~GeV and their tiny splitting must be explained.
Note that one can replace ${\bf L}=L_e-L_{\mu }-L_{\tau }$ symmetry by other symmetry, which will give approximate
${\bf L}$. For this purpose the anomalous ${\cal U}(1)$ symmetry of string origin is a good candidate \cite{Shafi:2000su}-\cite{Kitabayashi:2000nq}.
In our scenario  the charged lepton sector also plays an important role. In particular, the structure (\ref{YE})
is crucial for the predictions presented in the previous sections. We wish to understand this structure also by symmetry principles.
For this a non-Abelian discrete flavor symmetries can be very useful \cite{Pakvasa:1977in}-\cite{Mohapatra:2005wg}.
Therefore, in addition, we introduce $S_3$ permutation symmetry. $S_3$ will be broken in two steps:
$S_3\to S_2\to 1$. Since in the neutrino sector we wish to have $S_2$ symmetry, we will arrange for that sector
to feel only the first stage of breaking.

Thus, the model we present here is based on $S_3\tm {\cal U}(1)$ flavor symmetry. The $S_3$ permutation group has
three irreducible representations ${\bf 1},~{\bf 1'}$ and ${\bf 2}$, where ${\bf 1'}$ is an odd singlet while
${\bf 1}$ and ${\bf 2}$ are true singlet and doublet respectively.
With doublets denoted by two component vectors, it is useful to give the product rule
\vs{-0.3cm}
\beq
\begin{array}{cc}
 & {\begin{array}{cc}
\hs{-1.8cm}&
\end{array}}\\ \vspace{0.5mm}
\begin{array}{c}
 \\
 \end{array}\!\!\!\!\!\hs{-0.2cm} &{\left(\begin{array}{c}
x_1
\\
 \vs{-0.4cm}
\\
x_2
\\
\end{array}\right)_{\bf 2}\tm }~
\end{array}
\hs{-0.75cm}
\begin{array}{cc}
 & {\begin{array}{cc}
\hs{-1.8cm}&
\end{array}}\\ \vspace{0.5mm}
\begin{array}{c}
 \\
 \end{array}\!\!\!\!\!\hs{-0.2cm} &{\left(\begin{array}{c}

y_1
 \vs{-0.5cm}
\\

\\
y_2
\\
\vs{-0.5cm}
\\
\end{array}\right)_{\bf 2}=\l x_1y_1+x_2y_2\r_{\bf 1}~\oplus  ~\l x_1y_2-x_2y_1\r_{\bf 1'}~\oplus  ~}~
\end{array}
\hs{-0.75cm}
\begin{array}{cc}
 & {\begin{array}{cc}
\hs{-1.8cm}&
\end{array}}\\ \vspace{0.5mm}
\begin{array}{c}
 \\
 \end{array}\!\!\!\!\!\hs{-0.2cm} &{\left(\begin{array}{c}

x_1y_2+x_2y_1
 \vs{-0.5cm}
\\

\\
x_1y_1-x_2y_2
\\
\vs{-0.5cm}
\\
\end{array}\right)_{\bf 2} }~
\end{array}
\label{S3product}
\eeq
where subscripts denote the representation of the corresponding combination. The other products are very
simple. For instance ${\bf 1}\tm {\bf 1}={\bf 1}$, ${\bf 1'}\tm {\bf 1}={\bf 1'}$, etc.

 As far as the ${\cal U}(1)$ symmetry is concerned, a superfield $\phi_i$ transforms as
\beq
{\cal U}(1):~~~\phi_i\to e^{iQ_i}\phi_i~,
\la{trans-u1-ZN}
\eeq
where $Q_i$ is the ${\cal U}(1)$ charge of $\phi_i$.
The ${\cal U}(1)$ symmetry will turn out to be anomalous.
Such an anomalous $U(1)$ factor can appear in effective field theories from string theory upon compactification
to four dimensions.  The apparent anomaly in this ${\cal U}(1)$ is canceled
through the Green-Schwarz mechanism \cite{Green:1984sg}. Due to the
anomaly, a Fayet-Iliopoulos term $-\xi \int d^4\te V_A$ is always generated \cite{Witten:1981nf,Fischler:1981zk} and the corresponding
$D_A$-term  has the form \cite{Dine:1987xk}-\cite{Dine:1987gj}
\beq
\fr{g_A^2}{8}D_A^2=\fr{g_A^2}{8}\l -\xi +\sum Q_i|\phi_i|^2\r^2~,~~~\xi =\fr{g_A^2M_P^2}{192\pi^2}{\rm Tr}Q~.
\la{FI-D-A}
\eeq
In SUSY limit one chiral superfield should acquire a VEV in order to set $D_A$-term to be zero.

For $S_3\tm {\cal U}(1)$ breaking we introduce the MSSM singlet scalar superfields $\vec{S}, \vec{T}, X$, $Y$ and $Z$.
(vector symbols will denote $S_3$ doublets). We also introduce a discrete ${\cal Z}_4$ R-symmetry under which the superfields
transform as $\phi_i \to e^{i\fr{\pi }{2}Q_Z}\phi_i$ and the superpotential changes sign: $W\to -W$.
The transformation properties - the $S_3$ `membership', ${\cal U}(1)$ and ${\cal Z}_4$ charges - of all involved
 superfields are given in Table \ref{t:1}.
%
%%  begin table 1
%
\begin{table} \caption{Transformation properties under $S_3\tm {\cal U}(1)$, and $Q_Z$-charges of ${\cal Z}_4$ parity:
$\phi_i \to e^{i\fr{\pi }{2}Q_Z}\phi_i$, $W\to -W$.
 }

\label{t:1} $$\begin{array}{|c||c|c|c|c|c|c|c|c|c|c|c|c|}

\hline
\vs{-0.3cm}
 &  &  &  &  &  &  &  & & & &&\\

\vs{-0.4cm}

& ~\vec{S}~& ~\vec{T}~&~ Y~ &~Z ~& ~e^c_1 ~& ~ \vec{e}^{\hs{0.1cm}c}~&~l_1~&~ \vec{l}~  & ~N_1 ~ & ~N_2~ &~h_u~& ~h_d~\\

&  &  &  &  &  &  &  &  & &&&\\

\hline

\vs{-0.3cm}
 &  &  &  &  &  &  &  & & &&&\\

~S_3~& {\bf 2} & {\bf 2} & {\bf 1} & {\bf 1} & {\bf 1}& {\bf 2} & {\bf 1} & {\bf 2} &{\bf 1} & {\bf 1} & {\bf 1} & {\bf 1} \\

\vs{-0.3cm}
&  &  &  &  &  &  &  & & &&& \\

\hline
\vs{-0.3cm}
 &  &  &  &  &  &  &  & &&&&\\
\vs{-0.3cm}
~{\cal U}(1)~& 0 & -1 &2 &0 &4 &2 &1 &0 &-1 &k &0&-2\\

%\vs{-0.3cm}
&  &  &  &  &  &  &  &  & &&&\\

\hline

\vs{-0.3cm}
 &  &  &  &  &  &  &  & &&&&\\
\vs{-0.3cm}
~Q_Z~& 2 & 0 &2 &2 &-1 &-1 &-1 &-1 &-1 &1 &2&2\\

%\vs{-0.3cm}
&  &  &  &  &  &  &  &  & &&&\\

\hline

\end{array}$$

\end{table}
%
%%   end table 1
%
%

Let us first discuss the symmetry breaking. The most general  renormalizable `scalar' superpotential consistent with symmetries
has the form
\beq
W_{sc}=Y\vec{T}^2+\fr{\lam_1}{2}Z(\vec{S}^2-\La^2)+\fr{\lam_2}{3}\vec{S}^3+\fr{\lam_3}{3}Z^3~.
\la{Wsc}
\eeq
From the $F$-flatness conditions $F_{\vec{S}}=F_{\vec{T}}=F_Z=0$ we have the solutions
$$
\lan \vec{S}\ran =\l 0, ~V_S\r ~, ~~~~~~{\rm with}~~~~~~V_S=\La \l 1+2\lam_2^2\lam_3/\lam_1^3\r^{-1/2}
$$
\beq
\lan Z\ran =V_S\lam_2/\lam_1~,~~~~\lan Y\ran =0~ .
\la{VEVconfig}
\eeq
From $F_Y=0$ we get the condition $\lan \vec{T}^2\ran =T_1^2+T_2^2=0$ which is satisfied by
$\lan \vec{T}\ran =V_T\cdot \l 1, ~i\r $ with unfixed $V_T$ from the superpotential.
However, the non--zero value of $V_T$ is insured from the cancelation of $D_A$-term of (\ref{FI-D-A}).
Namely, with $\xi <0$, we have $V_T =\sq{-\xi/2}$.
Thus, all VEVs are fixed already in the unbroken SUSY limit and there are no flat directions.
We need to make sure that VEV configurations remain also stable with inclusion of higher order operators.
At the renormalizable level there are no couplings between $\vec{T}$ and $\vec{S}$. However, higher order interactions between
these states may change the winding of their VEVs. The lowest order   operators of this type respecting all other symmetries
are $\fr{1}{M_{\rm Pl}^2}Y\vec{T}^2(\vec{S}^2+\vec{S}Z+Z^2)$. In the presence of these couplings the $F_Y=0$ condition gets modified and we
obtain
\beq
\lan \vec{T}\ran =V_T\cdot \l 1, ~i(1+\eta )\r ~,~~~{\rm with }~~~\eta \sim \ep_s^2~,~~~~{\rm where}~~~~~\ep_s=V_S/M_{\rm Pl}
\la{VEVconfig1}
\eeq
(with $\lan Z\ran \sim V_S$).
As we see, the winding of $\lan \vec{T}\ran $ is slightly changed (with $\eta \ll 1$, i.e. $\ep_s\ll 1$).
This change will not have any impact in the charged lepton sector and for the light neutrino masses and mixings. However,
this will turn out to be important  in shifting the right--handed neutrino mass degeneracy and therefore for resonant leptogenesis.

We will use the following parametrization
\beq
\fr{\lan Z\ran }{M_{\rm Pl}}\sim \fr{V_S}{M_{\rm Pl}}\equiv \ep_s ~,~~~~~~\fr{V_T}{M_{\rm Pl}}\equiv \ep ~.
\la{eps}
\eeq
These two parameters will be used in expressing hierarchies between masses and mixings of the leptons.
All non-renormalizable operators that we consider in the charged lepton sector will be cut off by appropriate powers of
the Planck scale $M_{\rm Pl}$ and therefore in those operators the powers of $\ep_s$ and $\ep $ will appear.
There are also operators cut off with a different scale  which can be obtained by integrating out some vector-like states
with masses below $M_{\rm Pl}$.

\begin{figure}[t]
\begin{center}
\hs{-0.3cm}
\resizebox{0.8\textwidth}{!}{
  \vs{-3cm}\includegraphics{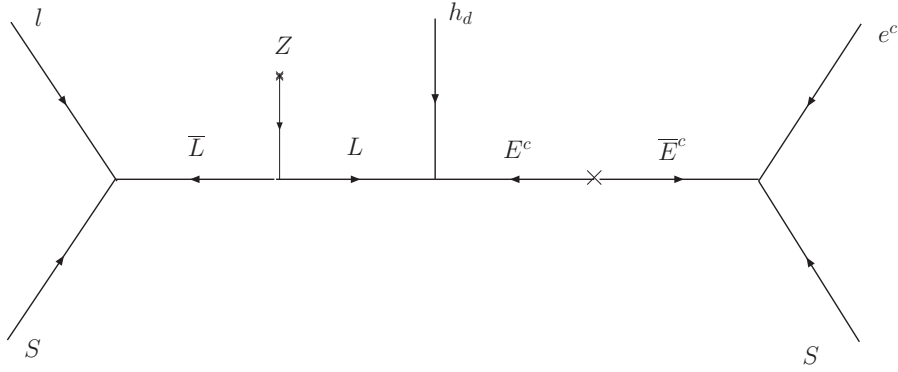}
}
%\put(-50,-10){$\lan Z\ran $}
%\vspace{5cm}       % Give the correct figure height in cm
\vs{-13cm}
\caption{Diagram generating the operator of Eq. (\ref{tau-op})}
\label{fig:4}       % Give a unique label
\end{center}
\end{figure}

Let us start with the charged lepton sector. For the tau lepton mass the operator
\beq
\fr{1}{\lan Z\ran M_*}\l \vec{l}\cdot \vec{S}\r_{\bf 1}\l \vec{e}^{\hs{0.15cm}c}\cdot \vec{S}\r_{\bf 1}h_d~,
\la{tau-op}
\eeq
is relevant, where $S_3$ contraction is in the singlet ${\bf 1}$-channel. This operator  can emerge by decoupling of heavy $L$, $E^c$ states
 in ${\bf 1}$ representation of $S_3$, as shown in the
diagram of Fig. \ref{fig:4}. Eq. (\ref{tau-op}) gives $\lam_{\tau }\sim V_S/M_*$, where $M_*$ is a mass
of $E^c, \ov{E}^c$ states.

Next, we include the following Planck scale suppressed operators:
\beq
\fr{1}{M_{\rm Pl}} \vec{l}\cdot \vec{e}^{\hs{0.15cm}c}\cdot (\vec{S}+Z)h_d +
\fr{1}{M_{\rm Pl}^2}l_1  \vec{e}^{\hs{0.15cm}c}\cdot \vec{T}\cdot (\vec{S}+Z)h_d +
\fr{1}{M_{\rm Pl}^3}e_1^c  \vec{l}\cdot \vec{T}^2\cdot (\vec{S}+Z)h_d +
\fr{1}{M_{\rm Pl}^4}l_1e_1^c\vec{T}^3\cdot (\vec{S}+Z)h_d~.
\la{charged-Ops}
\eeq
Substituting appropriate VEVs in (\ref{tau-op}), (\ref{charged-Ops}) and taking  into account that
$\vec{l}=(l_2 ,~l_3), ~\vec{e}^{\hs{0.15cm}c}=(e_2^c ,~e_3^c)$,
for the charged lepton Yukawa matrix we obtain
\begin{equation}
\begin{array}{ccc}
 & {\begin{array}{ccc}
 e^c_1& ~e^c_2&~e^c_3~
\end{array}}\\ \vspace{1mm}

\begin{array}{c}
l_1~\\ l_2~ \\l_3~
 \end{array}\!\!\!\!\!\hs{-0.1cm} \hs{0cm}&{\left(\begin{array}{ccc}

 \ep_s\ep^3& \ep_s\ep &\ep_s\ep
\\
\ep_s\ep^2& \ep_s &0
 \\
 \ep_s\ep^2&0 & \lam_{\tau }
\end{array}\right)}~,
\end{array}  \!\!  ~~~~~
\label{YE-derived}
\end{equation}
which nearly has the desired structure of (\ref{YE}).
It is easy to see that the $(1,1)$ entry of (\ref{YE-derived}) not presented in  (\ref{YE}) does not change  the predictive relations obtained in sect. \ref{sec:imprmodel}.
The $(1,3)$ and $(3,1)$ entries have no new parameters, and they shift relations in (\ref{YE}) by $\stackrel{<}{_\sim }1\%$.
Indeed, from (\ref{YE-derived}) we conclude that
\beq
\ep \simeq 0.13-0.2~,~~~~~~\ep_s\sim \lam_{\tau }\ep^2~
\la{ep-values}
\eeq
(this provides $\lam_e:\lam_{\mu }:\lam_{\tau }\sim \ep^5:\ep^2:1$, which is
compatible with the observed hierarchies).
Therefore,  the results of  sect. \ref{sec:imprmodel} are robust.

Now we turn to the neutrino sector. With transformation properties given in
Table \ref{t:1}, and for  integer $k>0$, the relevant couplings have the form
\beq
\begin{array}{cc}
 & {\begin{array}{cc}
\hs{-1.8cm}N_1\hs{0.2cm} &\hs{0.3cm} N_2
\end{array}}\\ \vspace{0.5mm}
\begin{array}{c}
l_1\vs{0.1cm}\\~ \vs{-0.3cm} \\ \vec{l}
 \end{array}\!\!\!\!\!\hs{-0.2cm} &{\left(\begin{array}{cc}
\fr{Z}{M_{\rm Pl}}  &~\fr{\vec{T}^{k+1}}{M_{\rm Pl}^{k+1}}
 \vs{-0.5cm}
\\
&
\\
0 &~\fr{\vec{T}^k}{M'M_{\rm Pl}^{k-1}}

\end{array}\right) h_u}~,~~~~~~~
\end{array}
\begin{array}{cc}
 & {\begin{array}{cc}
\hs{-1cm}N_1\hs{0.9cm} &\hs{0.9cm} N_2
\end{array}}\\ \vspace{1mm}
\begin{array}{c}
N_1 \\ N_2
 \end{array}\!\!\!\!\!\hs{-0.2cm} &{\left(\begin{array}{cc}

 0&~~~(Z\hs{-0.1cm}+\hs{-0.1cm}\vec{S})\fr{\vec{T}^{k-1}}{M_{\rm Pl}^{k}}
\\
 ~(Z\hs{-0.1cm}+\hs{-0.1cm}\vec{S})\fr{\vec{T}^{k-1}}{M_{\rm Pl}^{k}}&~\fr{\vec{T}^{2k}}{M_{\rm Pl}^{2k}}
\end{array}\right)M_R}~.
\end{array}  \!\!
\label{Ynu-MN-S3model}
\eeq
$M'$ is some cut off scale lower than $M_{\rm Pl}$.
%We will assume that $V_T/M'\sim 1$.
%
%Discussing the VEVs of the powers of $\vec{T}$, we should take into account the properties of
%$\lan \vec{T}\ran $'s configuration given in (\ref{VEVconfig1}). One can easily check out that
%\beq
%\begin{array}{cc}
%&\\
%\fr{1}{M_{\rm Pl}^2}\lan \vec{T}^2\ran \propto ~\ep^2\ep_s^2\cdot {\bf 1}~+\hs{-0.8cm}
%\end{array}
%\begin{array}{cc}
% & {\begin{array}{cc}
%\hs{-3cm}&
%\end{array}}\\ \vspace{0.5mm}
%\begin{array}{c}
% \\
% \end{array}\!\!\!\!\!\hs{-2cm} &{\ep^2\cdot \left(\begin{array}{c}
%i
%\\
% \vs{-0.4cm}
%\\
%1
%\\
%\end{array}\right)_{\bf 2}}~,~~~~~~~
%\end{array}
%
%\begin{array}{cc}
%&\\
%\fr{1}{M_{\rm Pl}^3}\lan \vec{T}^3\ran \propto ~\ep^3\cdot {\bf 1}~+\hs{-0.8cm}
%\end{array}
%\begin{array}{cc}
% & {\begin{array}{cc}
%\hs{-3cm}&
%\end{array}}\\ \vspace{0.5mm}
%\begin{array}{c}
% \\
% \end{array}\!\!\!\!\!\hs{-2cm} &{\ep^3\ep_s^2\cdot \left(\begin{array}{c}
%1
%\\
% \vs{-0.4cm}
%\\
%i
%\\
%\end{array}\right)_{\bf 2}}~,~
%\end{array}
%\label{Tk-prop}
%\eeq
%and so on. These configurations are useful for investigation of Dirac and Majorana matrices.
We have found one interesting
example: for $k=5$, and $\fr{V_T}{M'}\sim 1$
we obtain
\beq
\begin{array}{ccc}
 & {\begin{array}{cc}
\hs{-1.8cm} &
\end{array}}\\ \vspace{0.5mm}
\begin{array}{c}
 \\ \\
 \end{array}\!\!\!\!\!\hs{-0.2cm} Y_{\nu }=\hs{-0.3cm}&{\left(\begin{array}{cc}

 \ep_s &~\ep^6
 \vs{-0.5cm}
\\
&
\\
0&~\ep^4
 \vs{-0.5cm}
 \\
 &
\\
 0& ~i\ep^4

\end{array}\right) }~,~~~~~~~~~~~~
\end{array}
\begin{array}{cc}
 & {\begin{array}{cc}
\hs{-1.8cm} &
\end{array}}\\ \vspace{0.5mm}
\begin{array}{c}
 \\ \\
 \end{array}\!\!\!\!\!\hs{-0.2cm} M_N=&{\left(\begin{array}{cc}

 0 &~\ep_s\ep^4
 \vs{-0.5cm}
\\
&
\\
\ep_s\ep^4&~\ep_s^2\ep^{10}

\end{array}\right)M_R}~
\end{array}
\label{Ynu-MN-UZmodel}
\eeq
(where we have used the property $\lan \vec{T}\ran^{10}/M_{\rm Pl}^{10}\sim \ep_s^2\ep^{10}$).
Making a rotation of $N_{1,2}$ states to set (1,2) entry of the first matrix of (\ref{Ynu-MN-UZmodel}) to zero and at the same time performing
phase redefinitions we will arrive at the form of (\ref{Ynu-MN}) with
\beq
M=M_R \ep_s\ep^4\sim M_R\ep^6\lam_{\tau }~,~~~~~\al \sim \ep_s ~,~~~~ \bt \sim \ep^4~,~~~
\left |\de_N^*+\de_N^{\hs{0.3mm}'}\right |\sim \ep^6\ep_s\sim \ep^8\lam_{\tau }~,
\la{params}
\eeq
and
\beq
\sq{|\De m_{\rm atm}^2|}\sim \fr{\lan h_u\ran^2}{M_R}~,~~~~~~~~~
\left |\de_{\nu }^*+\de_{\nu }^{\hs{0.3mm}'}\right |\sim \fr{\ep^2}{\sq{2}}~.
\la{deltas-S3}
\eeq
Therefore, we get the right magnitude for $\De m_{\rm sol}^2/\De m_{\rm atm}^2$, while experimentally measured value
of $|\De m_{\rm atm}^2|$ dictates $M_R=\l 10^{13}-10^{14}\r $~GeV.
Thus, from (\ref{params}) we can estimate the absolute value of the RHN mass.
For $\tan \bt \simeq 2$ (the value used for numerical studies in sect. \ref{sec:leptogenesis}) we get $M=(10^6-10^8)$~GeV.
This range includes the values of RHN masses such that SUSY gravitino problem is avoided.
At the same time we get $\left |\de_N^*+\de_N^{\hs{0.3mm}'}\right |\sim 10^{-9}-5\cdot 10^{-8}$ and $x\sim \lam_{\tau }/\ep^2\sim 1$.
All these values work well for resonant leptogenesis (see Fig. \ref{fig:3}).
Therefore, we conclude that the model presented in this section works well for inverted neutrino mass hierarchical
scenario and also insures the success of resonant leptogenesis.

Finally, we briefly comment on some aspects of low energy phenomenology of the presented model.
The superpotential  term $Yh_uh_d$ is allowed by symmetries of the model, which has a potential for generating the
$\mu $-term with $\lan Y\ran \sim 1$~TeV induced after SUSY breaking. As far as the quark
sector is concerned, the Yukawa couplings $qu^ch_u$ and $qd^ch_d$ are allowed with the following ${\cal U}(1)$
and ${\cal Z}_4$ charge assignment: $Q(q, u^c, d^c)=(y, -y, 2-y)$ and $Q_Z(q, u^c, d^c)=(1, -1, 1)$.
This charge assignment is flavor independent. However, if desired one can also select flavor dependent
charges for understanding of hierarchies between quark masses and CKM mixings.
The freedom in the selection of $y$  can be exploited for
the simultaneous cancelation of  $SU(3)^2\tm {\cal U}(1)$ and $SU(2)^2\tm {\cal U}(1)$
mixed anomalies via Green-Schwarz mechanism (achieved with $y=7/9$).
Also, one can verify that
$SU(3)^2\tm {\cal Z}_4$ and $SU(2)^2\tm {\cal Z}_4$  anomalis are automatically zero  with the above
${\cal Z}_4$ charge assignment. Therefore, ${\cal Z}_4$ can be identified as a discrete gauge symmetry.
One remarkable feature is that with the ${\cal Z}_4$ symmetry, matter parity is automatic. Indeed, with the $Q_Z$
charge assignment, a $Z_2$ subgroup of ${\cal Z}_4$ R-symmetry remains unbroken.
Therefore, ${\cal Z}_4$-symmetry insures that the model has realistic phenomenology with a stable LSP.

\section{Conclusions}

In this paper we have presented a new class of models which realizes an inverted spectrum for neutrino
masses. These models predict a definite correlation between neutrino mixing angles $\te_{12}$ and $\te_{13}$.
Deviation of $\te_{12}$ from $\pi /4$  is controlled by the value of $\te_{13}$. Our results are given in
Eqs. (\ref{pred-12-23})-(\ref{range-mee}) and plotted in Figs. \ref{fig1}, \ref{fig:2}.

We have presented a concrete model based on an $S_3$ permutation symmetry augmented with a discrete ${\cal Z}_4$ R-symmetry and
 ${\cal U}(1)$ symmetry acting on the three flavors.

Our model can naturally lead to resonant leptogenesis since two right--handed neutrinos are quasi-degenerate.
The predictions of our model are testable in forthcoming experiments.

\vs{0.5cm}

\hs{-0.7cm}{\bf Acknowledgments}
\vs{0.2cm}

\hs{-0.7cm}We thank S. Gabriel and A. Pilaftsis for useful discussions and E. Lisi for allowing us to use
their figure from Ref. \cite{Fogli:2005cq}.
The work is supported in part by DOE grants DE-FG002-04ER41306 and DE-FG02-04ER46140.

\bibliographystyle{unsrt}

\end{document}